\documentclass[12pt,letterpaper]{article}

\usepackage{graphicx} 
\begin{document}

\title{\vskip1cm
Study of transport properties in rippled channels
\vskip.5cm}
\author{Ivan Herrera\\ 
Instituto de F\'{\i}sica y Matem\'aticas, \\
Universidad Michoacana \\
58060 Morelia, Michoac\'an, M\'exico\\
E-mail:iherrera@ifm.umich.mx\\
\bigskip
\and 
Alberto Mendoza, Eduardo S. Tututi\footnote{Corresponding author: E.S. Tututi. Fax +52 443 3167257}\\ 
Facultad  de Ciencias  F\'{\i}sico-Matem\'aticas, \\
Universidad Michoacana \\
58060 Morelia, Michoac\'an, M\'exico\\
E-mail: almend@zeus.umich.mx, tututi@zeus.umich.mx
}
\date{}
\maketitle
\begin{abstract}
We study   dynamical properties of bouncing particles  inside of   channels with sinusoidal walls. Taking as parameters the amplitude and the phases
between the walls we study the transmitivity and its dependence on these parameters. We find an analytical approximation for transmitivity for small amplitudes, which is corroborated by numerical calculation.   
\end{abstract}

{\bf Pacs numbers:}  05.45.Pq, 73.23.Ad, 73.63.Kv 

\newpage
\section{Introduction}
Two dimensional billiards are systems consisting of a free point particle
moving in a two dimensional region bounded by rigid walls.
The billiards problem  leads to  both classical and quantum
interesting and well studied class of Hamiltonian systems such as the stadium 
\cite{stadium} and the Sinai billiards \cite{sinai}. Basically, from a 
classical point of view the problem consists in to find the trajectory of a 
particle which bounces elastically on the boundary walls.
From the quantum  point of view, the problem consists in  solving the 
Helmholtz  equation by requiring that the wave function vanishes on the 
boundary. At the mesoscopic level, a  question of great interest in the transport properties in open billiards, such as quantum dots is the influence of the shape of the boundary. In fact, it has been observed that coherent backscattering and 
conductance fluctuations are strongly shape dependent \cite{berry}.
Due to the development of nanostructure devices \cite{marcus,burki} (in addition to the quantum dots we can mention the quantum wires),  the transport properties  with both regular and irregular profiles is of current interest \cite{nakamura,kouwenhoven}. As matter of fact it has been observed that  when the nanotubes are elongated they present a pattern of regularity in the shape. Many of these systems have been studied and  show regular and chaotic dynamics \cite{luna1}. Quantum and classical billiards systems, have been helpful in the understanding the quantum manifestation of classical chaos and quantum transport for large values of  energy. At low temperatures, of order of few mK, the mean free path of electrons is much larger than the size of the quantum devices, diffusive effects disappear and the electrons show a rather ballistic behavior. In this case a classical analysis is enough for studying the transport properties in these quantum and mesoscopic devices.

In this work we study a two dimensional rippled  open channel. The profile of walls of the channel are sine-like but, in general, with different phases. The case of one rippled wall has been previously studied in Ref. \cite{luna1,luna2}.
We investigate the dependence  of the transport properties on the difference of phases as well as the dependence on the amplitude between the walls in the ballistic regime, we obtain an analytical approximation for reflexivity for small ripples. We observe that the chaotic regime appears faster when the difference of the phase between the walls is near to $\pi$ than when they are near to be in phase for a given amplitude. Our results are corroborated by numerical calculations.

\section{The rippled channel}
\begin{figure}[ht]
\begin{center}
\includegraphics[width=12cm]{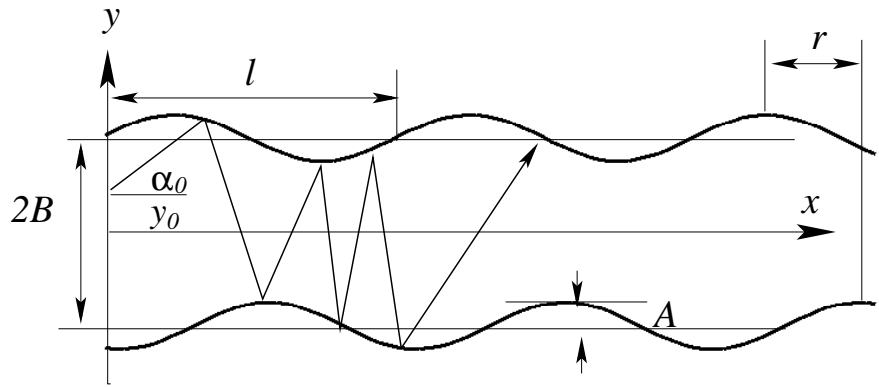}
\caption{\label{figu1}The profiles of the upper  and the lower walls correpond to a
difference of phase $r$}
\end{center}
\end{figure}

Let us consider an open channel with rough walls such as the shown in Fig. \ref{figu1}. 
The upper and lower walls are determined, respectively, by
\begin{eqnarray}
y_1 & = & b + a \sin 2 \pi x
\nonumber\\
y_2 & = & -b +a\sin 2\pi\left( x+r\right) 
\end{eqnarray} 
where, instead of the physical size, we  use dimensionless variables $x=\frac{X}{l}$, $y=\frac{Y}{l}$, 
$b=\frac{B}{l}$, $a=\frac{A}{l}$, $L={L^{\prime}\over l}$ with $l$ being the length of the period and $L^{\prime}$  the length of the channel,
$A,B$ are respectively the separation between the walls and the amplitude. The variable $r$ denotes the difference of phase between the upper and lower walls and it is restricted to the values $\mid r\mid \leq \frac{1}{2}$.

The trajectory of each particle bouncing elastically inside the channel is define by the initial condition 
($y_0$, $\alpha_0 $), where $y_0$ denotes the high at which the particle was initially dropped and $\alpha_0 $ the initial angle. Since the particles bounces elastically, it satisfies that the angle of incidence equals to the reflection angle,
meseared from the tangent line to the curve. For our work  an important transport property is the transmitivity, $T$, that is defined by the flux of the transmitted particles divided by the incoming flux. To compute this numerically we dropped  $10^4$ particles  from the left side of the channel (at $x=\frac{1}{2}$), of length $L=2$, by 10 sources uniformly distributed along the $y$ axis (each source eject 1000 particles). We assume that the particles do not interact among themselves and 
that no multiple collision with the walls occurs before the particle leaves away the corresponding wall.  

The angular distribution (for each source) of the dropped particles is 
\begin{eqnarray}
\rho(\alpha_0)=\frac{n_0}{2B}\cos(\alpha_0),
\end{eqnarray}
where $\alpha_0$ is measured with respect to $x$ axis  as is shown in Fig. \ref{figu1} and $n_0$ is the total number of particles.

\section{Poincar\'e sections}
Let us  discuss a channel whose walls are separated away  by a distance $2b=0.2$. 
In order to  obtain   all the different types of trajectories in 
the phase space ($x_n$,$p_n$), were $x_n$ and $p_n$ are respectively the value of the $x$ coordinate at the $n$-th collision with 
the upper wall and $p_n$  the $x$ component of the momentum after the $n$-th collision.  We  throw  particles with different initial conditions in the phase space. 
The $x$ component of the momentum is normalized by the velocity.
The system can be carry out in practice  if we identify the particles as electrons and resort to the  ballistic regime which is realized at low temperatures in
nanostructures where the size of the sample  $L^{\prime}$ is such  that $\lambda_e<<L^{\prime}<L_e$,
where $\lambda_e$ is the electron wavelength and $L_e$ is the mean free path \cite{nakamura}.   

\begin{figure}[ht]
\begin{center}
\includegraphics[width=8cm]{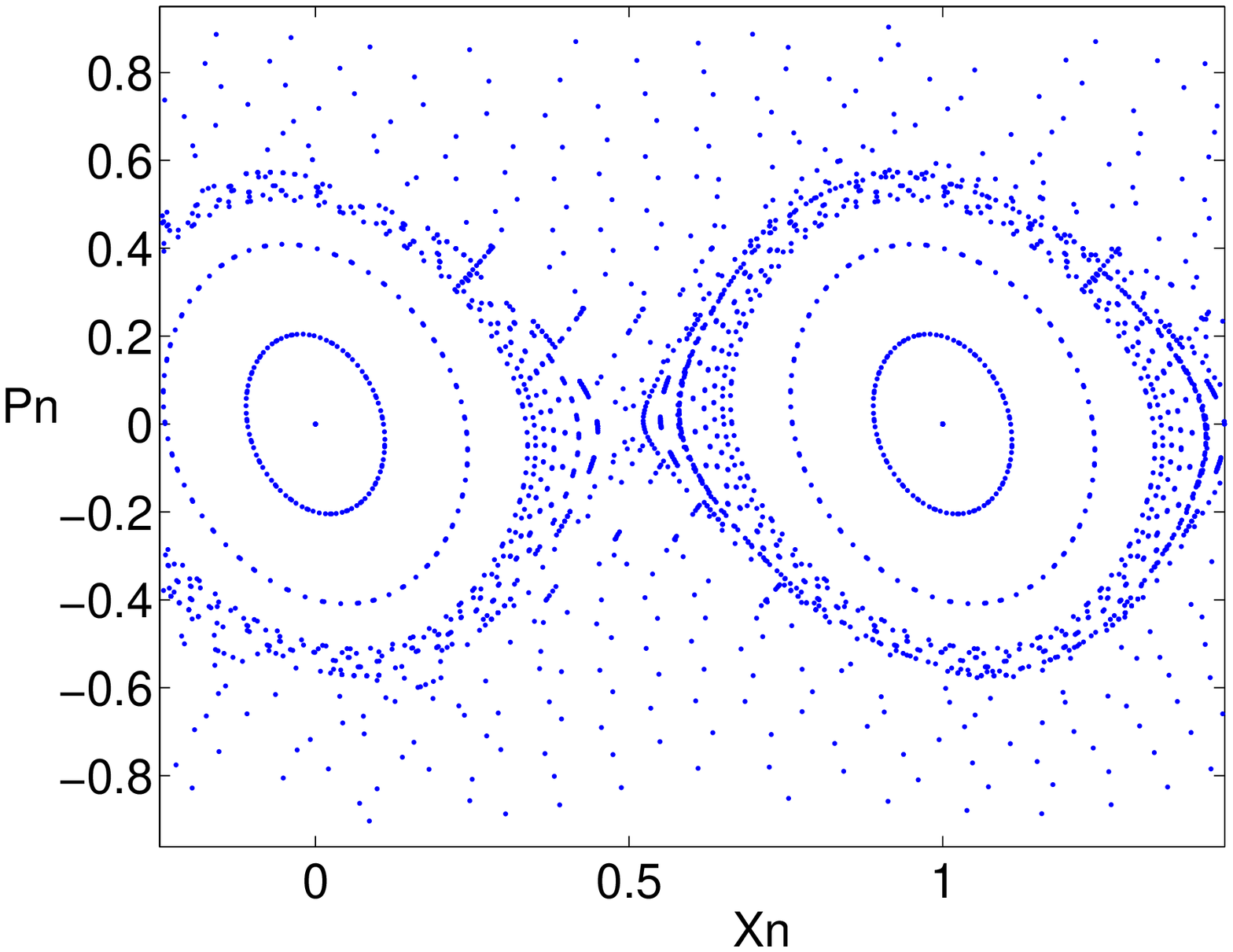}
\caption{\label{figu3} Poincar\'e plot for $b=0.1$, $a=0.01$ and $r=\frac{1}{2}$.}
\end{center}
\end{figure}

\begin{figure}[ht]
\begin{center}
\includegraphics[width=8cm]{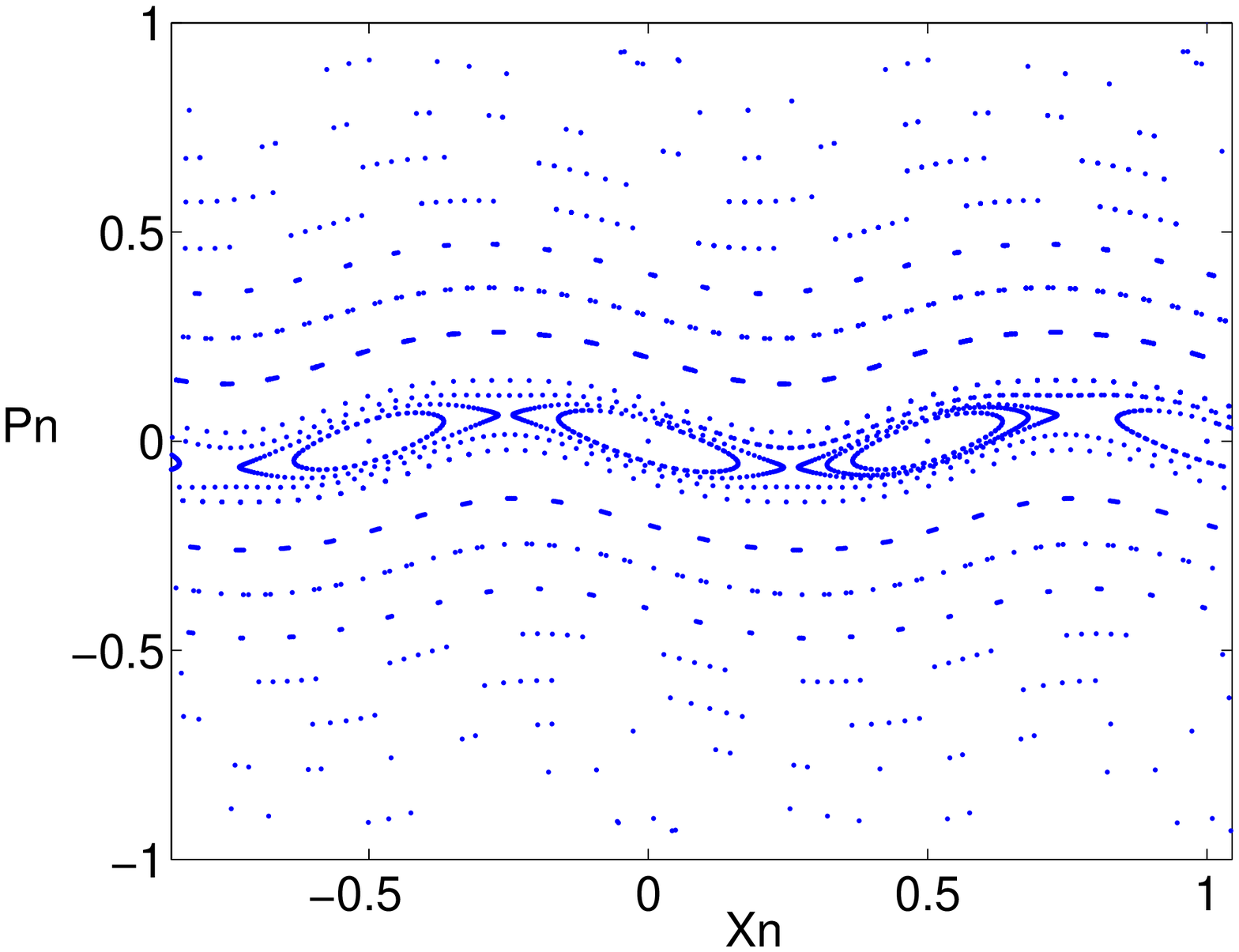}
\caption{\label{figu4}Poincar\'e plot for $b=0.1$, $a=0.01$ and $r=0$.}
\end{center}
\end{figure}

\begin{figure}[ht]
\begin{center}
\includegraphics[width=8cm]{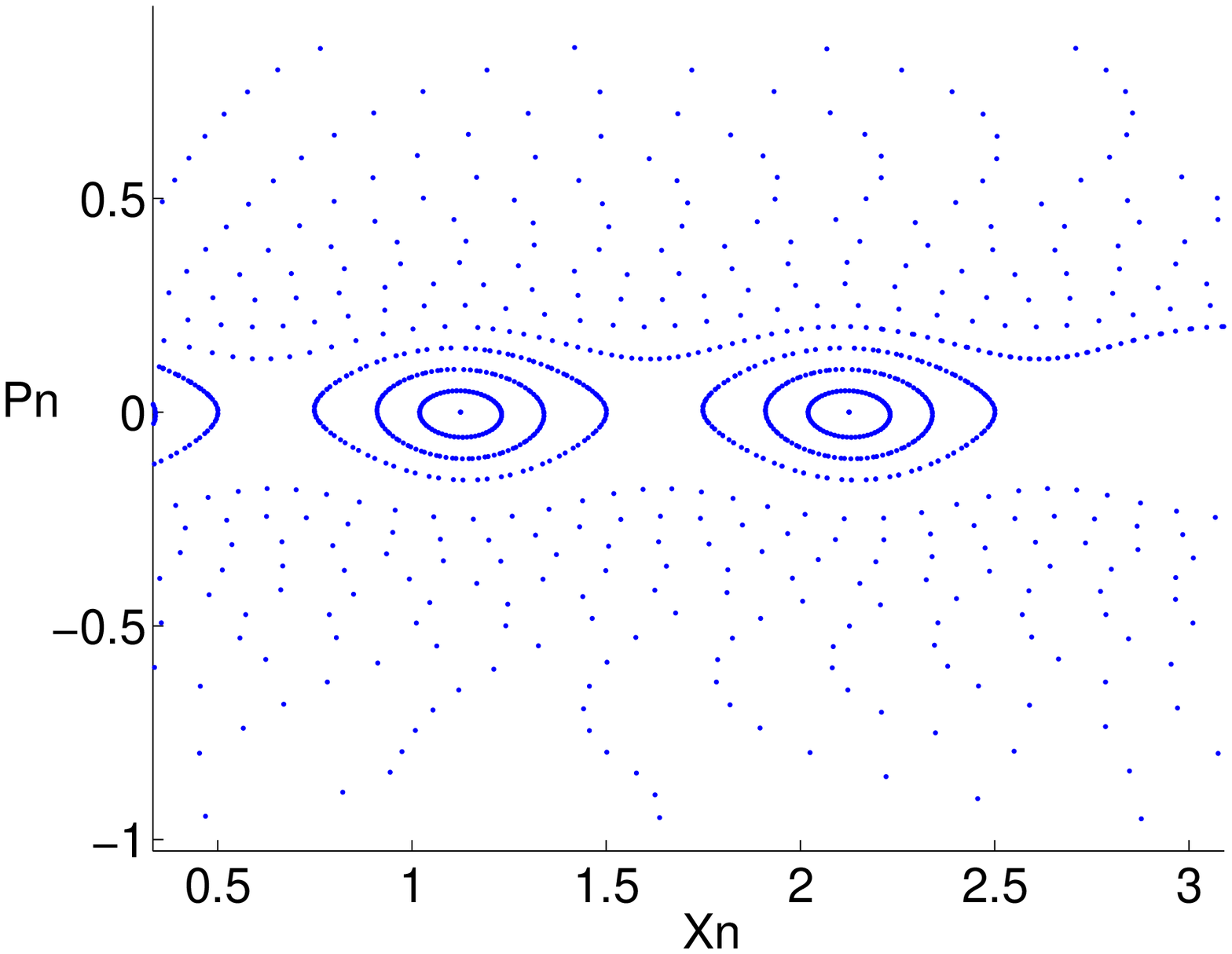}
\caption{\label{figu5}Poincar\'e plot for $b=0.1$, $a=0.001$ and $r=\frac{1}{4}$.}
\end{center}
 \end{figure}

\begin{figure}[ht]
\begin{center}
\includegraphics[width=8cm]{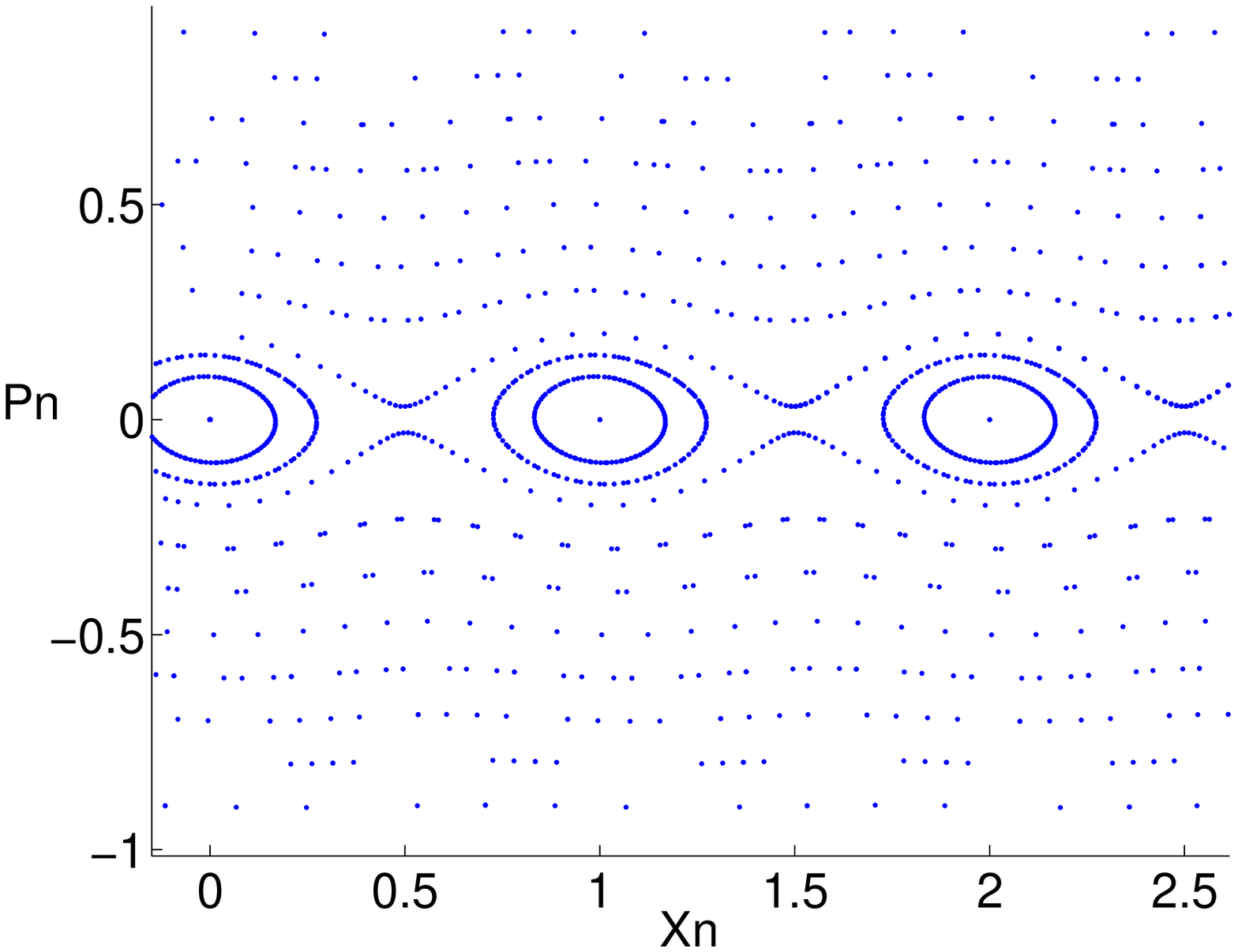}
\caption{\label{figu7}Poincar\'e plot for $b=0.1$, $a=0.001$ and $r=\frac{1}{2}$.}
\end{center}
\end{figure}

\begin{figure}[ht]
\begin{center}
\includegraphics[width=8cm]{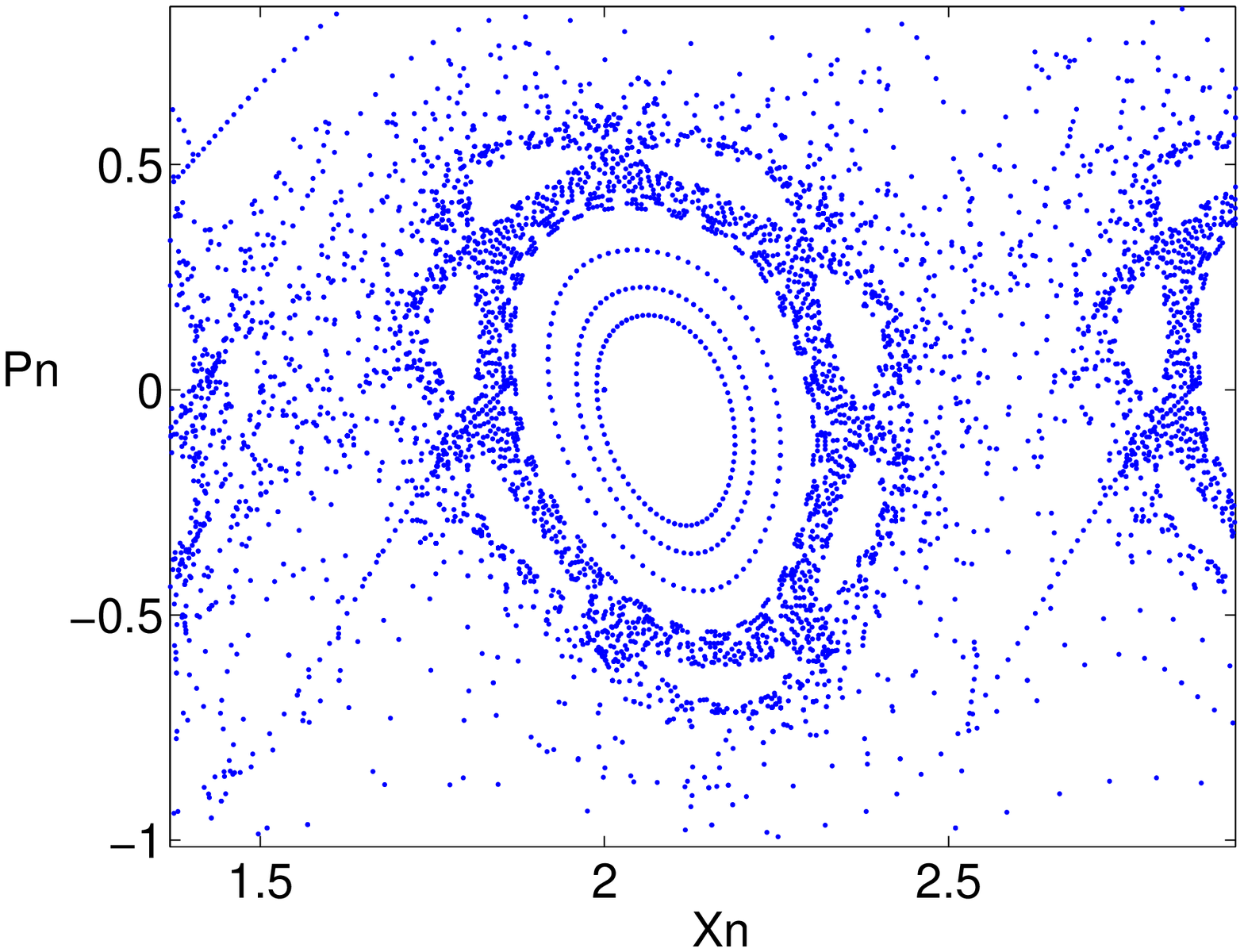}
\caption{\label{figu8}Poincar\'e plot for $b=0.1$, $a=0.02$ and $r=\frac{1}{3}$.}
\end{center}
\end{figure}

Some Poincar\'e sections are shown in Figs. \ref{figu3}-\ref{figu8}. Notice that
Poincar\'e sections are moved along  the $x$ direction by a quantity $\frac{1}{4}$. 
For small values 
of $a$ the corresponding Poincar\'e sections are similar to a pendulum phase space (Figs. \ref{figu5} and \ref{figu7}).
When we go to higher amplitudes, the separatrix in $x_n\approx 0.5$ of Fig. \ref{figu3} becomes chaotic. For higher values  of $a$
we get a region of trapped particles (ellipses or deformed ellipses) surrounded by a chaotic sea (Fig. \ref{figu8}), this occurs when the last KAM curve breaks and then all chaotic regions are connected. 
For small values of $a$, the elliptic orbit indicates that the particles are 
trapped around the center of the ellipse (not all of these points are fixed points). Notice that for a fixed point $Y_n=Y_{n+1}$, where $Y_n$ is the trajectory of the particle after the $n$-th collition. This leads to the condition $X_0=\frac{1}{2\pi}{\rm Arcos}\left( \frac{r}{4\pi ab}\right)$, where $X_0$ is a fixed point, so in order to have a fixed point the phase $r$ must satisfies the relation $|r|\leq 4\pi ba$. 

When the amplitude is small, the trapped particles execute librational motion (similar to the pendulum oscillations around the stable equilibrium point), they move adiabatically forward and backward. The only change due to the phase, for small amplitudes, is that the center of the ellipses are moved in the $x$ direction but they maintain almost constant in the $p$ direction, near to $p_n=0$ (see  Figs. \ref{figu5} and \ref{figu7}).

From our results, we observe that for values of the phase near to $r=\frac{1}{2}$, whenever the amplitude is increased, the chaotic regime appears faster than for  values  near to $r=0$. This behavior is observed as well for the last KAM curves; as we go near to $r=\frac{1}{2}$ the last KAM curves break sooner. 
As an example when the walls are in phase the separatrix becomes chaotic until $a=0.04$  and the last KAM curve breaks until $a=0.075$. For $r=1/3$ the last KAM curve breaks around $a=0.016$ and for  greater values, for instance $a=0.02$ (Fig \ref{figu7}) it appears a cluster of seven elliptical curves surrounding the principal first order resonance island. When $r=1/2$ the chaotic behavior appears around $a=0.005$ and the last KAM curve breaks around $a=0.015$. There is an appreciable difference between region of trapped particles when the profiles are in phase than when $r=0.5$.
For the same amplitude $a$, we can see these differences from Figs. \ref{figu3} and 
\ref{figu4} there are more trapped particles when the phase is $r=0.5$ than when the phase is $r=0$.

\section{Transport properties for small amplitudes}

\begin{figure}[ht]
\begin{center}
\includegraphics[width=8cm]{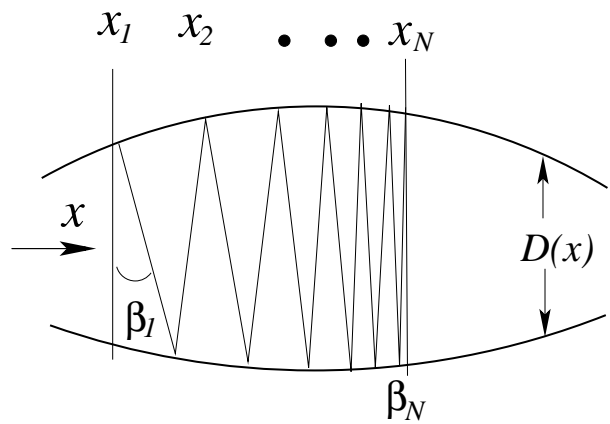}
\caption{\label{label}Figure to illustrate the adiabatic approximation.}
\end{center}
\end{figure}

For small amplitudes (e.g. $a=0.001$)  and when the direction, at which the particles are dropped, is almost parallel to the vertical,  the particles collide almost perpendicularly with the walls, this means that $|\dot{y}|>>|\dot{x}|$. As the particles bounces elastically  the energy  is conserved. If, in addition, we use the principle of adiabatic invariance we can show that the motion in the $x$-direction is described approximately by \cite{percival}:
\begin{eqnarray}
(\dot{x})^2={\frac{2E}{m}}-\left( {\frac{C}{D(x)}}\right)^2,
\label{eq3}
\end{eqnarray}   
where $C$ is a constant. In this case  $C=\dot{y}D(x)$ is the action, $E$ is  the total energy 
and $D(x)$ is the separation betwen the walls at the $x$ point, $v\approx \dot{y}$. 
The term $V(x)=\left( {\frac{C}{D(x)}}\right)^2$ can be interpreted as an effective potential. If $\frac{a}{b}<<1$, $V(x)$ becomes
\begin{equation}
V(x)\approx\frac{C^2}{4b^2}\left( 1-\frac{a}{b}\left({\rm sen}(2\pi x)-{\rm sen}(2\pi (x+r))\right)\right).
\end{equation}
For the case of $r=\frac{1}{2}$, $V(x)$ takes the form of pendulum-like potential and explain why the phase space in this instance  has a pendulum-like form 
(see Fig. \ref{figu7}).
We may label, in Eq. (\ref{eq3}), the variable $x$ with the index $n$ to indicate the $n$-th collition at the $x$ point. Let $\dot{x}_{n}=v\,{\rm cos}(\alpha_n)$ and   $\dot{x}_{m}=v\,{\rm cos}(\alpha_m)$ be the $x$ velocity at the $n$-th collition and $m$-th collition respectively. From Eq. (\ref{eq3}) we can obtain a relation between $\alpha_n$ and $\alpha_m$:
\begin{eqnarray}
v^2({\rm cos}^2\alpha_n-{\rm cos}^2\alpha_m)=\left( \frac{C}{D_m(x)}\right)^2- \left( \frac{C}{D_n(x)}\right)^2,
\label{eq4}
\end{eqnarray}
where $D_n(x)\equiv D(x_n)$.
Let be $\beta_n=\frac{\pi}{2}-\alpha_n$,  the angle that the trajectory makes with the vertical at the $n$-th collition so  that $\beta_n$ is small and it becomes small and small as it reaches the turning point $x_N$ (see Fig. \ref{label}), at this point $\beta_N\approx 0$. Let be $\beta_c$ the critical angle at which the particle  was dropped in $x=\frac{1}{2}$ and does not turn back. The trajectory corresponding to this initial condition is associated with the the largest amplitude of librational motion at the center of the ellipses. Let us define  $x_{Nm}$ as the point (being bigger than $x=\frac{1}{2}$) for which the greatest ellipse crosses the $x$ axis.
 
Let $\beta_n=\beta_c$, $D_n=D_c=D(\frac{1}{2})$ and $\beta_m=\beta_N$, $D_m=D(x_{Nm})$ substituting these into Eq. (\ref{eq4}) and keeping only terms of the order of $a$ and $\beta_c^2$ we obtain:
 \begin{eqnarray}
(\beta_c)^2\approx \frac{a}{b}({\rm sen}(2\pi(x_{Nm}+r))-{\rm sen}(2\pi x_{Nm})+
{\rm sen}(2\pi r)).
\label{aprx}
\end{eqnarray}    
 If $r=\frac{1}{2}$, then $x_{Nm}\approx 0.75$ and:
 \begin{eqnarray}
 \beta_c^2\approx\frac{2a}{b}.
\end{eqnarray} 
This result is similar to the semiplane case, the only difference is that in this case  the amplitude of the channel is twice bigger than the semiplane case \cite{luna1}. Notice also for $r=0$ and $x_{Nm}=0.75$ and from Eq. (\ref{aprx})
that $\beta_c$ is of order of zero. Here, we can see  why the great difference between $r=0$ and $r=0.5$, even for small amplitudes.
If $r=\frac{1}{4}$, it can be seen from Fig. \ref{figu5} that $x_{Nm}=\frac{7}{8}$ 
and then $\beta_c^2\approx(\sqrt{2}+1)\frac{a}{b}$, this means that for  the case of $r=\frac{1}{4}$ the channel has a higher value of resistivity than for case of $r=\frac{1}{2}$.
 
For small amplitudes, we see numerically that $\beta_c$  is almost the same for different values of $y_0$  where $y_0$, is the initial point for which the particle was dropped, this means that $\beta_c$ depends mainly on $r$, $b$ 
and $a$. We have seen numerically that for small amplitudes the phase space is dominated by KAM 
tori in a pendulum-like manner, the motion of the particles with angles $\alpha_0\approx\frac{\pi}{2}$ can be considered in the adiabatic approximation, 
thus the velocity in the $x$ direction $v_x$, can be estimated as $v_x\propto\beta_0$ and produce a backward flux, such that the reflexivity, $R$,
is given 
\begin{eqnarray}
R=1-T\approx \frac{8B}{\pi n_0}\int_{0}^{\beta_c}\beta_0\rho(\frac{\pi}{2}-\beta_0)d\beta_0.
\label{eq9a}
\end{eqnarray}     
Using the approximation for $\beta_c$ given in  (\ref{eq4}), Eq. (\ref{eq9a}) becomes
  \begin{eqnarray}
R&\approx&\frac{4}{3\pi}\beta_c^3
\nonumber\\
&\approx& G(r)a^{\frac{3}{2}},
\label{eq10a}
\end{eqnarray}
where the coefficient 
\begin{equation}
G(r)=\frac{4}{3\pi b^{\frac{3}{2}}}\left( {\rm sen}(2\pi(x_{Nm}+r))-{\rm sen}(2\pi x_{Nm})+{\rm sen}
(2\pi r)\right)^{\frac{3}{2}}.
\end{equation}  
Notice also  from  equation (\ref{eq10a})  that for all phases the reflexivity is proportional to $a^{\frac{3}{2}}$ and the information of the phase is contained in the coefficient G because we have seen numerically that for small amplitudes $x_{Nm}$ varies very slow when $a$ is varied, this means that $x_{Nm}$ depends mainly on $r$. An important measurable quantity is the resistivity $\rho_R$ of the channel that is related with the reflexivity through  Landauer's formula \cite{landauer}.
\begin{eqnarray}
\rho\propto\frac{R}{T}
\label{resis}
\end{eqnarray}    
Since $T\rightarrow1$ as $R\rightarrow0$ and according to Eqs (\ref{eq10a}) and (\ref{resis}), we conclude that $\rho_R$ increases proportionally to $a^{\frac{3}{2}}$ independently of the phase $r$.

\section{Numerical results for reflexivity and transmitivity}
\begin{figure}[ht]
\begin{center}
\includegraphics[width=8cm]{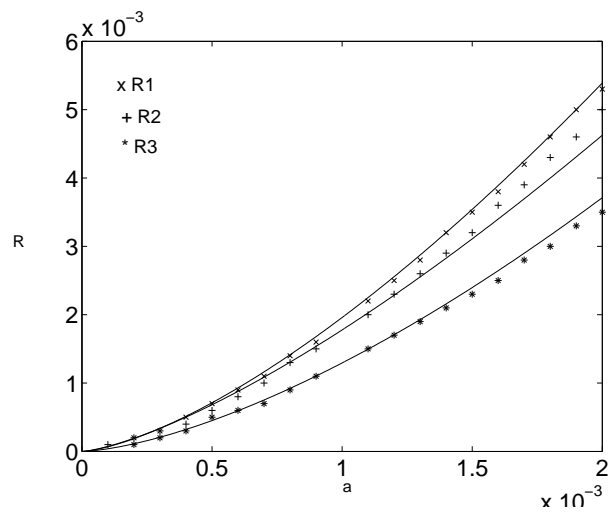}
\caption{\label{figu11} Reflexivity $R$ against amplitud $a$ for $r=\frac{1}{3}$ ($R_1$), for $r=\frac{1}{4}$ ($R_2$), for $r=\frac{1}{2}$ ($R_3$)}
\end{center}
\end{figure}

Fig. \ref{figu11} shows reflexivity against amplitude for three different phases $r=\frac{1}{3}$, $r=\frac{1}{4}$ and $r=\frac{1}{2}$ which correspond to the following numerical equations respectively $R_1=47.46a^{1.46}$, $R_2=24.53a^{1.38}$, $R_3=46.9a^{1.52}$. Our analysis predicts $R\propto a^{\frac{3}{2}}$ (Eq. \ref{eq10a}) which gives a good  agreement between the analytical 
and numerical results in powers of $a$. The analytical coefficients $G(r)$ for $r=\frac{1}{2}$ and $r=\frac{1}{4}$ are respectively $37.9$ and $49.7$, the discrepancies in  the coefficients, between the analytical and numerical results arise from  the  exponential grow of the numerical error  since we made a fitting data as a function of a power law. 

\begin{figure}[ht]
\begin{center}
\includegraphics[width=8cm]{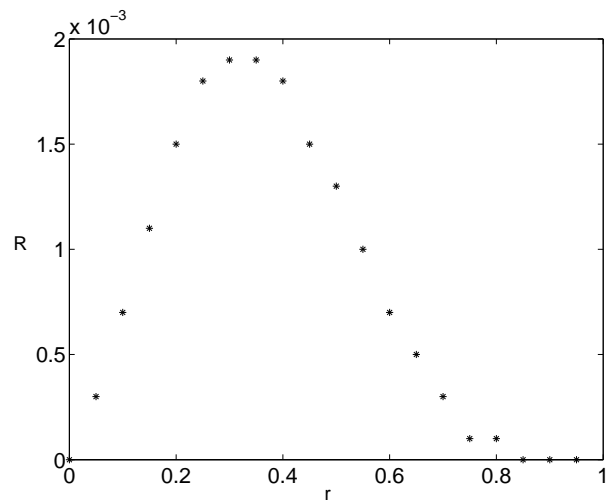}
\caption{\label{figu12} Numerical values for Reflexivity $R$ against phase $r$ and amplitude $a=0.001$. Notice that for the values $r=0$ and for $r=1$ it is same physical situation.} 
\end{center}
\end{figure}

In Fig. \ref{figu12} we show numerical values for resistivity against the  phase 
keeping constant the amplitude. We can see  that for $r=0$ the reflexivity 
vanishes; a rather expected result since from Eq. (\ref{aprx}) we obtain that $\beta_c\approx 0$. Moreover when the phase ($r$) is increased the reflexivity increases until it reaches its  maximun value at approximately $r=0.35$, then the reflexivity goes down to the value 0 at $r=1$,  which is the same geometrical situation that for $r=0$. Notice  that there were significant change in reflexivity only when a change on the phase $r$ is more than $\delta r=0.05$.

We have seen numerically that $T=1$ when $r=0$ if $a<0.005$, this does not mean that all the particles are transmitted, but the critical angle is so small that makes the computer insensitive to this values of angles, this implies $\beta_c<5\times 10^{-3}$, using Eq. (\ref{eq10a}) we see that  $R<1.25\times 10^{-7}$ this is in agreement with Eq (\ref{aprx}) that predicts $B_c\approx0$, which implies $R\approx0$. We can from  Eq. (\ref{resis})  that $\rho_R < 10^{-7}$.  In general the numerical calculations are in agreement with the analytical results.

\section{Conclusions}

We have analyzed a two dimensional channel with  rough walls. We studied the Poincar\'e sections for different values of  phase between the walls $r$ and the amplitude $a$. We found that Poincar\'e section are pendulum-like for small values of $a$. We also note that when the amplitude is increased
the chaotic regime appears sooner when we are near to $r=\frac{1}{2}$ than when we are near to $r=0$. Based on analytical and numerical results we found that resistivity 
$\rho \propto a^{\frac{3}{2}}$ is independent of the parameter $r$, such that the information about the phase is complete contained  in the proportionality coefficient $G(r)$. Our numerical results for small amplitudes, show that the system with the biggest reflexivity is around  $r=3.5$ and the system with the smaller reflexivity is when the walls are in phase and resistivity is of order of $a^2$. We have shown that the difference of phases between the walls play an important role in dynamic transport properties even for small values of the amplitude. Finally we would like to mention that our results could be useful in the analysis of transport properties in  elongated nanotubes  in the ballistic regime. 

\section*{Acknowledgments}
This work has been partially supported by  CIC-UMSNH.


\end{document}